\documentclass[12pt]{article}

\usepackage{setspace,graphicx,epstopdf,amsmath,amsfonts,amssymb,amsthm,versionPO}
\usepackage{marginnote,datetime,enumitem,rotating,fancyvrb}
\usepackage{hyperref,float}
\usepackage[longnamesfirst]{natbib}
\usdate

\excludeversion{notes}		
\includeversion{links}          

\iflinks{}{\hypersetup{draft=true}}

\ifnotes{%
\usepackage[margin=1in,paperwidth=10in,right=2.5in]{geometry}%
\usepackage[textwidth=1.4in,shadow,colorinlistoftodos]{todonotes}%
}{%
\usepackage[margin=1in]{geometry}%
\usepackage[disable]{todonotes}%
}

\makeatletter\let\chapter\@undefined\makeatother 

\usepackage{indentfirst} 
\usepackage{jfe}          

\usepackage{subcaption}
\usepackage{graphicx}
\usepackage{multicol}
\begin{document}

\setlist{noitemsep}  

\title{Dealer Strategies in Agent-Based Models}

\author{Wladimir Ostrovsky\thanks{Author's Address: Department of Economics, 
University of Kiel, Olshausenstr. 40, 24118 Kiel,
Germany, Email: w.ostrovsky@economics.uni-kiel.de.}\\
  CAU Kiel\\
  }

\date{10.12.2023}              


\renewcommand{\thefootnote}{\fnsymbol{footnote}}

\singlespacing

\maketitle

\vspace{-.2in}
\begin{abstract}

This paper explores the utility of agent-based simulations in realistically 
modelling market structures and sheds light on the nuances of optimal dealer 
strategies. It underscores the contrast between conclusions drawn from 
probabilistic modelling and agent-based simulations, but also highlights the 
importance of employing a realistic test bed to analyse intricate dynamics.
This is achieved by extending the agent-based model for auction markets by 
\cite{Chiarella.2008} to include liquidity providers. By constantly and 
passively quoting, the dealers influence their own wealth but also have 
ramifications on the market as a whole and the other participating agents. 
Through synthetic market simulations, the optimal behaviour of different dealer 
strategies and their consequences on market dynamics are examined. 
The analysis reveals that dealers exhibiting greater risk aversion tend to 
yield better performance outcomes. The choice of quote sizes by dealers is 
strategy-dependent: one strategy demonstrates enhanced performance with 
larger quote sizes, whereas the other strategy show a better results with smaller 
ones. Increasing quote size shows positive influence on the market in terms 
of volatility and kurtosis with both dealer strategies. However, the impact 
stemming from larger risk aversion is mixed. While one of the dealer strategies 
shows no discernible effect, the other strategy results in mixed outcomes, 
encompassing both positive and negative effects. 

\end{abstract}

\medskip



\thispagestyle{empty}

\clearpage

\onehalfspacing
\setcounter{footnote}{0}
\renewcommand{\thefootnote}{\arabic{footnote}}
\setcounter{page}{1}

\section{Introduction}

%
%

%

This paper develops an agent-based model centred around a limit orderbook 
(hereafter LOB) as a mainstay for information dissemination among market 
participants, facilitating opinion formation and subsequent price discovery.
It emphasises the introduction of a passive dealer who acts as a liquidity 
provider. In contemporary financial markets, auctions\footnote{This holds in 
particular for stock markets.} are the prevailing operational mode, where 
buyers and sellers simultaneously set their quotes. This results in complex 
dynamics that have yet to be understood from both a behavioural finance and 
economic perspective. In the subsequent sections, the model will be used as a 
test bed to analyse the dealer's optimal behaviour but also the impact on 
market dynamics. 

The presented work is primarily concerned with the introduction of a liquidity 
provider which is an essential part of financial markets. The impact of such 
agents has garnered attention from various perspectives. While numerous studies 
suggest that high frequency trading can disrupt price discovery, increase 
volatility beyond fundamentals, there is an equal body of research asserting 
that liquidity has expanded, as measured by increased order book depth and 
size.\footnote{See for example \cite{Froot.1992}, 
\cite{Zhang.2010}, \cite{AndreiA.Kirilenko.2011}, \cite{Cvitanic.2010} and 
\cite{Hasbrouck.2013} among others.} 
Additionally, spreads have significantly narrowed resulting in less room for 
manipulation and predatory strategies.\footnote{Refer to \cite{Vuorenmaa.2013} 
for a comprehensive literature review on these studies.} Given the complexity 
of financial markets and growing influence of computer-driven traders, it is 
important to further investigate the influence of liquidity providers.

This work primarily relates to \cite{Vuorenmaa.2014} where the authors utilise 
the classical agent-based model developed in \cite{Chiarella.2002} and 
\cite{Chiarella.2008}. \citeauthor{Vuorenmaa.2014}'s 
(\citeyear{Vuorenmaa.2014}) goal is to study high frequency traders (HFT; 
throughout the paper, also referred to as dealer) more realistically in 
situations of flash crashes. As in \cite{Avellaneda.2008}, HFTs optimise 
their inventory risk while providing continuous liquidity. The study 
incorporates an institutional trader who is able to initiate a flash crash 
through specific types of selling strategies. They find that the probability of 
a flash crash increases with more HFT agents or smaller inventory size. Market 
volatility is negatively correlated with the aforementioned factors and 
deriving optimality conditions without knowledge of the decision 
maker's\footnote{Can be thought as market participants or regulatory bodies.} 
preferences is unfeasible.

This model builds upon previous work of \cite{Chiarella.2002} and 
\cite{Chiarella.2008}. The papers introduce an order-driven market with 
multiple types of representative agents including fundamentalists, chartists 
and noise traders. The later extension of \cite{Chiarella.2008} allows agents 
to submit orders by maximising their expected utility and apply varying time 
horizons for different types of strategies. In addition, a passive liquidity 
provider is introduced to the model following \cite{Avellaneda.2008} where a 
strategy is developed for stock dealers who provide liquidity in a LOB. 
The authors combine the utility framework of \cite{Ho.1981} with results of 
microstructure literature on the limit orderbook. Following a two-step 
procedure the dealer determines an indifference price for a given asset by 
taking his inventory position as well as risk aversion into account. 
Afterwards, bid-ask quotes are derived from the probability of order arrivals 
and risk aversion.

The aim of this work is to contribute to the literature on 
financial agent-based models of auction markets and the research on optimal 
dealer quoting. In particular I will contribute with a detailed study of the 
wealth implications and price discovery driven by dealer behaviour.

The first part serves as a prelude to subsequent topics concerning the main 
model and addresses the necessity of agent-based models by critically 
examining probabilistic simulations as in \cite{Avellaneda.2008}. 
Subsequently, I integrate a risk-averse dealer in a LOB framework enabling a 
more realistic modelling of price discovery and feedback loops. 
The study will analyse two distinct dealer strategies: 1) An agent 
that optimises profits and inventory risk akin to \cite{Avellaneda.2008} and 2) 
a dealer resembling the approach in \cite{Fushimi.2018} who uses a 
practitioners rule to manage inventory risk. Furthermore, I compare the 
strategies with a naive liquidity provider who simply provides fixed volume 
around the mid-price. Other agent types deployed choose their allocation in the 
risky instrument by maximising their utility under constant relative risk 
aversion. While this extension adds complications to the simulation by 
introducing wealth-dependent allocations, it enables a realistic assessment of 
the feedback loop between an agent's wealth and different parameter 
sensitivities. 

My second contribution is to shed light on the influence of preference 
parameters on the dealer's optimal behaviour. For this purpose, several 
parameters in the behavioural function of the dealer have to be calibrated in 
order to create realistic dynamics. Establishing a baseline parameter set is 
crucial as it ensures the dealer's profitability which is her primary 
motivation. Moreover, I present the endogenous feedback loop and its 
implications for price changes and wealth. By choosing parameter values the 
dealer influences the market dynamics which the dealer and other agents then 
react to in turn. Conclusions on the optimal behaviour of the dealer will be 
derived along with the implications on the market. Given the dealer's decisions 
are entangled with price dynamics but also personal wealth, it is important to 
clearly present dependencies.

The findings underscore the importance of agent-based simulations as a valuable 
tool for achieving more realistic market-like structures. It is 
shown that conclusions on the dealer strategy under probabilistic modelling may 
be contradictory to agent-based simulations. Further, several critical features 
such as time dependent effects of inventory and its management cannot be 
analysed in full depth without a realistic test bed. 
The study explores sensitivities with respect to the dealer's preferences to 
understand their impact on both the dealer's wealth and market dynamics.
An increase in the dealer's risk aversion is found to elevate market volatility 
while reducing kurtosis of market returns. Overall dealers gain more wealth by 
operating more risk averse. Adjusting the dealer's maximum order size leads to 
opposing outcomes as the optimising dealer benefits through capitalising on 
larger trades while the other experiences constant wealth loss.
Markets clearly favour larger quote size as visible in decreased volatility and 
kurtosis. Stylised agents, in particular fundamentalists and chartists, augment 
their wealth gains as max. order size increases. Decreasing the dealer's 
sensitivity to her inventory will lead to a less abrupt downward adjustment of 
quoted volume and shows a positive impact on her wealth.\footnote{Later this 
sensitivity will be introduced as inventory skew.} Similarly market return 
variance and kurtosis are dampened as order size is adjusted less drastically.

The paper is organised as follows: Section 2 provides an overview of the 
literature. In particular the models most closely related to this work will be 
discussed. Section 3 continues by demonstrating the difficulties and limits of 
a theoretical and probabilistic evaluation of the market maker's optimal 
quoting rule. In section 4, the main model is described and linked to the 
original literature presented earlier. Finally, section 5 presents the results 
and section 6 concludes.

\section{Literature Review}

Agent-based models have gained increasing attention over the last decades. Much 
of economics research condense human behaviour into structural models with 
(boundedly) rational agents aiming to characterise features in an easily 
interpretable way.\footnote{See \cite{LeBaron.2002}.} The findings in the area 
of behavioural finance have supported the approach which is taken by 
agent-based 
financial markets research.\footnote{The curious reader can find an 
introduction into the topic in \cite{Shefrin.2007}.} Using a simulation 
environment in which individual behaviour is explicitly defined and agents 
interact with each other, sophisticated macro features can be generated that 
replicate facts of financial markets. 
This part will serve as an introduction into the topic, highlighting the, for 
this work, important contributions. Note that the review is rather exhaustive 
which is required to have the background for the model deployed in the main 
part.
Models designed by \cite{Chiarella.2002} and \cite{Chiarella.2008} are the 
foundation for the work presented here. The authors studied how different 
trading strategies affect price discovery in an order driven market as well as 
the effect of market design features on liquidity.\footnote{These trading 
strategies have been discussed widely in models with simple market clearing 
mechanisms. In particular these are models that deploy Walrasian auctioneers 
for market clearing; \cite{LeBaron.2001} summarises different types of market 
clearing mechanisms in section 3.2.}

Like the majority of real world trading, the model centres around a so 
called limit orderbook which serves as a platform for price 
discovery. The LOB takes all incoming orders, stores them as 
decreasing/increasing sequence for buy/sell orders in price-time priority, 
and finally matches buyers and sellers when bids and asks cross.
In their model, agents decide sequentially about sending orders which have a 
finite lifetime $\tau$. A simple mechanism is put in place to remove the oldest 
$\tau$ orders: at every timestamp $t$, $\nu$ is drawn from a uniform 
distribution, if $\nu < \omega$, a constant, the oldest $\tau$ orders are 
cleared. The platform determines the price at timestamp $t$ as the last 
transaction price in case a transaction occurred, otherwise it is 
calculated as 
$p_t = (p_t^{bid, best} + p_t^{ask, best}) / 2$. 
If no bids (asks) are present, then the previous price is used. In this model 
prices are generally positive and can take only 
pre-specified values on a grid determined by the tick size $\Delta$.

\cite{Chiarella.2002} assume a representative agent 
that combines fundamentalist, chartist and noise trader 
behaviour\footnote{\cite{Hommes.2018b} surveys the work on dynamic 
heterogeneous agent models, where he introduces the widely used trading 
strategies of chartists, fundamentalists and noise traders.} to 
forecast returns denoted with $\hat r$ 

\begin{equation}
\label{eq:1}
\hat r^i_{t, t + \tau} = g^i_1 \frac{(p^f - p_t)}{p_t} + g^i_2 
\bar r_{L^i} + n_i \epsilon_t
\end{equation}

where $g^i_1 > 0$, $g^i_2$ and $n_i$ are weights assigned to fundamentalists, 
chartists and noise traders respectively.
The latter have zero intelligence, thus base their prediction on a Gaussian 
random variable, $\epsilon_t$. The fundamental price, $p^f$, is only known to 
fundamentalists, upon which they base their trading decision. In particular, 
they want to buy (sell) the security if the market price is below (above) the 
fundamental price. Chartists extrapolate recent trends and, hence, base their 
decision on past returns, $\bar{r}_{L^i}$, calculated as the average 
return over the interval $L^i$ sampled from a uniform distribution 
$\mathcal{U}(0, L^{max})$
\begin{equation}
\label{eq:2}
\bar r_{L^i} = \frac{1}{L^i} 
\sum_{j=1}^{L^i} ln 
\frac{p_{t-j}}{p_{t-j-1}} \text{.}
\end{equation}

At every timestamp $t$, an agent is chosen to submit an order with a price to 
the LOB which is determined as 
$\hat p_{t+\tau}^i = p_t \exp(\hat r^i_{t+\tau})$. 
If the agent forecasts a price increase she decides to buy a unit of the 
stock and vice versa. It is implicitly assumed in this formulation that the 
agents know the price history as well as the prevailing fundamental price.

In their findings, the authors conclude that all three types of agents are 
necessary to create realistic return dynamics: the chartist reaction function 
generates large price jumps and clustering of volatility while fundamentalists 
tend to reduce the excursions away from the fundamental price. 
Chartists reinforce the clustering as they experience higher submission rates 
of matched orders in more volatile markets. Similarly the bid-ask spread is 
correlated with trading volume as the LOB is increasingly depleted around the 
mid-price.

\cite{Chiarella.2008} extend the original version to allow agents to submit 
orders by maximising their expected utility and apply varying time horizons for 
the different types of strategies. Specifically, agents are assumed to maximise 
their utility of wealth, $u$, under constant absolute relative risk aversion 
(hereafter, CARA)


\begin{equation}
\label{eq:3}
\operatorname*{max}_{Z^i_{t + \tau}} \mathbb{E} 
\left[u(W^i_{t+\tau}, \gamma^i)\right]
\end{equation}
\begin{equation}
\label{eq:4}
u(W, \gamma) = -e^{-\gamma^i W^i_{t + \tau}}
\end{equation}
where $\gamma$ is the risk aversion of agent $i$. The wealth, $W$, of each agent
at timestamp $t$ is defined as the sum of the market value of their individual 
stock portfolio and cash 
\begin{equation}
\label{eq:5}
W_t^i = q_t^i p_t + C_t^i
\end{equation}
where $q_t^i \geq 0 $ and $C_t^i \geq 0 $ are respectively stock and cash 
positions. As implied by the inequalities, short-selling is not 
permitted. Given the CARA utility defined above, the agents can calculate their 
optimal allocation, $Z$, towards the risky asset by\footnote{Note that in this 
formulation, the focus is on determining the proportion of an agent's wealth 
allocated to the risky asset. This approach differs from the original one used 
by the authors, which was expressed in terms of the number of stocks. Here, 
the investment decision is based on the currency amount to be invested in 
the risky asset, which is a product of the wealth allocation $Z_t^i$ and the 
agent's total wealth, $W_t^i$. Consequently, $W_t^i$, is canceled out in the 
calculation.}

\begin{equation}
\label{eq:6}
Z_t^i = \frac{ln(\hat p_{t+\tau}^i / p_t)}{\gamma^i {\sigma_t^{2}} W^i_t}
\end{equation}
where the actual investment made is independent of the individual wealth as 
$W^i_t$ is cancelled out. The market volatility is denoted by $\sigma^{2}$.

In this extended model, the authors provide evidence that a significant 
chartists component is needed to emphasise trends and create jumps, which are 
typical of real financial markets. Further it is shown that all types of agents
are required to create a realistic shape of the LOB: fundamentalists reduce the 
imbalance between buy/sell orders stemming from noise trader behaviour and 
chartists widen the distribution away from mid-prices.

%
%
%
%
%

A strand of the operations research literature is concerned with
defining optimal rules on how dealers should place their quotes conditional on 
a set of objectives. \cite{Avellaneda.2008} define the problem in terms of an 
expected utility framework where the dealer optimises her inventory through 
setting bid-ask quotes under uncertainty from the transaction risk caused by 
the stochastic movement of the stock price and arrival rate of orders. The 
stock price is assumed to evolve according to a Brownian Motion without drift 
while the arrival of orders is modelled as a Poisson intensity.\footnote{This 
is especially related to the studies of \cite{Gabaix.2003} and 
\cite{Maslov.2001} who show that the size of market orders follows a power law 
distribution along with the work of \cite{Potters.2002} who explain that the 
price impact of market orders is proportional to the natural logarithm of order 
size.}

A two step process is derived where the dealer first estimates a reservation 
price, $p_t^r$, making her indifferent to trade
\begin{equation}
\label{eq:7}
p_t^r = p_t - q_t \gamma \sigma^2_t (T-t)
\end{equation}
where $q_t$ is the dealer's inventory at timestamp $t$, $\gamma$ defines her 
risk aversion and $\sigma^2_t$ is the variance of market returns. The 
reservation price is the difference between the current price and the inventory 
adjusted for risk aversion and variance. Hence, the reservation price is tilted 
away from the inventory side. In the subsequent step, the bid-ask spread, 
$\delta^a + \delta^b$, is computed around the 
reservation price
\begin{equation}
\label{eq:8}
\delta^a + \delta^b = \gamma \sigma^2_t (T-t) + \frac{2}{\gamma} ln(1 + 
\frac{\gamma}{\kappa})
\end{equation}
where $\kappa$ is the arrival intensity. As the intensity of order arrivals 
rises, dealers are willing to trade more frequently. Simultaneously, they 
reduce the spread, thereby decreasing the likelihood of holding an unbalanced 
inventory. Likewise the spread increases in risk aversion and variance, but is 
reduced closer to session end. The latter is used by the dealer to increase the 
probability to end the day flat in terms of her inventory and is reflected in 
the term $T-t$ with $T$ referring to the session's closing time. Probabilistic 
simulations were performed which show that the optimal strategy has a lower 
variance in the P\&L profile and final inventory compared to a naive benchmark 
strategy.

After the flash crash of 2010, academics began to investigate the impact of 
high-frequency traders on market stability. \cite{Vuorenmaa.2014} combine the 
work by \cite{Chiarella.2008} and \cite{Avellaneda.2008} described above to 
model HFT agents more realistically in situations of flash crashes. The authors 
introduce an institutional investor who uses an execution algorithm which 
follows a selling strategy that triggers a flash crash. They find that the 
probability of a flash crash rises with an increasing number of HFT agents or 
tighter inventory control. Contrary, market volatility will decrease with 
higher number of HFT agents. It was noted that setting variables optimally 
is not possible without knowing the preferences of the regulator/decision 
maker.

The here presented work will build on the models developed in the 
above-described contributions. In particular I am going to unite agent-based 
models from economics literature and theoretical results from operations 
research of optimising dealers to derive insights in an simulation environment.

\section{Inventory Risk in Liquidity Provision Strategies} \label{sec:Inventory 
Risk}

This section will review the baseline \cite{Avellaneda.2008} model (AS model, 
dealer or strategy) and demonstrate shortcomings when considering large order 
sizes or inventory risks. Further, this part deals as a prologue to the model 
derived in the next section by emphasising the importance of agent-based 
simulations.

\begin{figure}[!htbp]
\centerline{\includegraphics[width=5.5in]{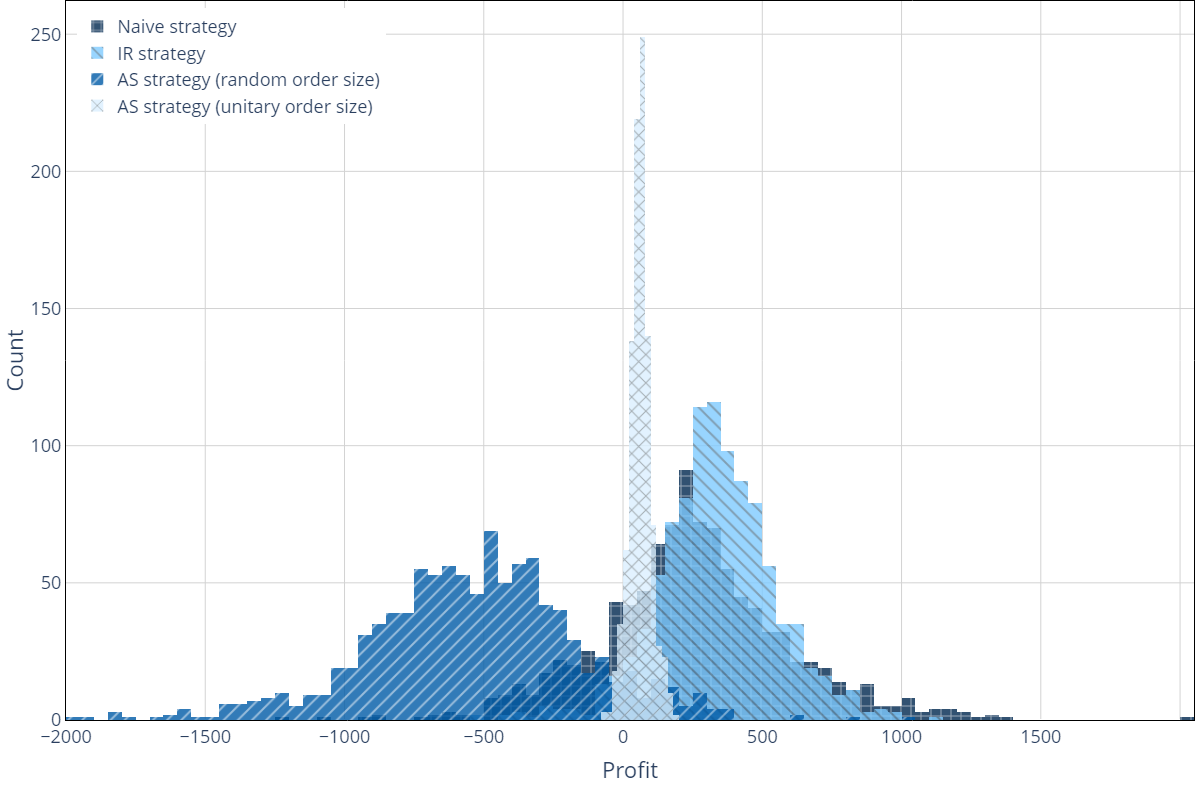}}

  \caption{Terminal wealth distributions for the AS model with 
  unitary (light grey) and random order size (light blue) as well as the IR 
  strategy 
  (blue) taking the inventory adjustment of \cite{Fushimi.2018} into account. 
  For a complete comparison a naive strategy (dark blue) is included. The 
  histograms contain 1000 simulation runs which are based on the same parameter 
  space as in \cite{Avellaneda.2008} if not stated differently.} 
  \label{fig:histogram_1}
\end{figure}

The AS model is simulated by iterating through time $T$ in $dt$ 
increments. At every timestamp $t$, the dealer computes bid and ask prices 
according to equation (\ref{eq:8}) based on the prevailing market price and 
her inventory. At the next timestamp, $t + dt$, the market price is updated by 
a random increment of $\pm \sigma \sqrt{dt}$. With a probability of 
$\lambda^a(\delta^a)dt$ the dealer's inventory decreases by one unit while her 
cash increases by $p_t + \delta^a$. Conversely, with a probability of 
$\lambda^b(\delta^b)dt$ the inventory increases by one unit while cash 
decreases by $p_t - \delta^b$. The probabilities $\lambda^a(\delta^a)$ and 
$\lambda^b(\delta^b)$ denote the execution intensity of the dealer's orders 
where each single term stands for the arrival of buy and sell orders 
respectively. Thus, a change in wealth is determined as the difference between execution price, derived from equations (\ref{eq:7}) and (\ref{eq:8}), and the prevailing market price, which follows a Brownian Motion. Figure \ref{fig:histogram_1} displays the distribution of terminal wealth.

Notably, the AS model is a probabilistic simulation where a single unit of the 
underlying is traded and the market price is independent from the actual 
trading - both assumptions are not realistic. To demonstrate the issue of simplification to an unitary order size, the simulation is adjusted by introducing a random variable 
$x_t$ drawn from a Gamma distribution, specifically $\Gamma(\alpha=2, 
\beta=15)$. At every timestamp $t$ where the dealer makes a 
transaction, $x_t$ determines the trade size; inventory as well as 
wealth are adjusted incorporating the random size. 

Figure \ref{fig:histogram_1} illustrates profit distributions for a range of 
strategies. The AS model is simulated following their work with a unitary order 
size. Additionally, the profit distribution of the adjusted AS model is shown when order 
size varies randomly, as described above. To draw a meaningful comparison a 
naive dealer strategy is created. The naive strategy quotes a constant order 
size of $\phi_t^{max} = 15$ around the market price. To support the underlying 
thesis, the spread is calculated the same way as in the AS model in equation 
(\ref{eq:8}). 

The terminal profit distribution clearly shows a profit decrease with the adjusted AS 
strategy as mean profits are visibly shifted to the left. This is a consequence of a larger inventory and thus a greater effect on wealth from market price fluctations. In contrast, the baseline strategy with unitary order sizes maintains a low inventory and as such allows the bid-ask spread revenue to largely neutralize the impact of price fluctuations. However, this compensatory mechanism is absent in the adjusted AS strategy. Counter to 
what would be expected, the naive strategy significantly outperforms the AS 
dealer. The reason can be found in the adjustment of the reference price.
The naive dealer uses a symmetric quote around the mid-price. This earns her a 
higher compensation from the bid-ask spread than the AS dealer who shifts her 
reservation price toward the inventory side, thereby losing out on potential 
profit.

In practice, risk managers often introduce value-at-risk limits which, simply 
put, translates into inventory size limits. To show how the wealth is impacted 
by a risk management rule, an inventory adjustment from 
\cite{Fushimi.2018} is used (hereafter referred to IR model, dealer or 
strategy). The authors extend the AS model by introducing a dynamic order 
sizing framework to rule out excessive inventory by placing smaller order sizes 
in the direction of the excess inventory
\begin{equation}
\label{eq:9}
\phi_t^{bid} =
\begin{cases}
	\phi_t^{bid, max} &\text{if } q_t \leq 0 \\
	\phi_t^{bid, max} e^{\eta^{bid} q_t} &\text{if } q_t > 0\\
\end{cases}
\end{equation}

\begin{equation}
\label{eq:10}
\phi_t^{ask} =
\begin{cases}
	\phi_t^{ask, max} &\text{if } q_t \geq 0 \\
	\phi_t^{ask, max} e^{-\eta^{ask} q_t} &\text{if } q_t < 
	0\\
\end{cases}
\end{equation}
where $\phi_t^{ask, max}$ and $\phi_t^{bid, max}$ are the maximum ask and bid 
order sizes the dealer is willing to quote at any point in time, $\eta^{ask}$ 
and $\eta^{bid}$ are the adjustment factors (also called inventory skew) that 
will decrease the quote size as soon as the inventory is different from 
zero.\footnote{Here, I deviate and ignore the calculation of  
a reservation price but set it to the market price, $p_t^r = p_t$. This will be 
picked up in the later sections as it does not proof useful to follow the 
two-step procedure.}
Simulating this strategy like the previous examples, a similarly good 
performance as with the naive strategy can be achieved. The higher mean is 
attributable to a larger quote size at the reversing transaction and, thus, 
greater compensation through the bid-ask spread while lower variance is due 
to a more constrained inventory position.

To motivate the subsequent part of this work, these examples 
highlight the general need for having deeper analysis within a simulation 
framework. The profit distribution in Figure \ref{fig:histogram_1} raises the 
obvious question why a naive strategy outperforms an optimised model. A simple 
rule based stock management shows superior profits. The absence of a realistic simulation platform renders it impractical to analyze the underlying behavior, especially when trying to determine the causes of specific instances of under-/overperformance. Most 
notably, the dealer impact on market dynamics remains unexplained. An 
accumulation of inventory was eliminated in the simulations as the 
probabilities will eventually be high enough so that a reverse transaction will 
take place pulling the inventory towards zero. Unfortunately, real markets do 
not strictly following any well known distribution; as such mean reverting 
behaviour to the zero inventory is not granted. In fact, without restricting 
the dealer, she may accumulate vast amounts of stocks. By introducing the 
inventory rule, the impact on price discovery is seemingly different. Take a 
situation where the dealer's inventory is saturated so that she removes one 
side of the quote. Considering the LOB's depth and the distance between orders, the calculated mid-price may experience jumps. Similarly, the AS dealer will set her 
quotes around an reservation price which can tilt the market price in a 
particular direction. These effects will move through the market by means 
of feedback loops with the other participants and eventually lead to a 
different price path. 
Further, the AS strategy assumes that the market returns follow a Brownian 
motion. However, prices may experience trends and perpetuating behaviour in 
real markets. To account for endogenous effects, market participants and their 
interactions have to be modelled explicitly. To understand these time-dependent 
effects enforced by inventory risk management and dynamic interaction, the need 
for an agent-based model is bolstered.

\section{The Model} \label{sec:Model}

I introduce a new model that improves on the results shown in the previous 
section. It is suitable as a test-bed to analyse optimal behaviour of the 
dealer and agents and can outline market reactions. In the following I will 
describe the different components of the model including the assets traded, 
agents who interact with each other through a limit orderbook and the 
simulation procedure.

\subsection{Assets} \label{sec:Assets}
For simplicity, assume a single tradable asset in the form of an arbitrary 
stock. The price of the stock is determined through the interaction of the 
different agents via a LOB. Furthermore, agents exchange cash and stock in 
their individual transactions. Cash is assumed to carry zero interest and to 
have a constant currency value. 
While the stock price is determined by agent interactions, fundamental type 
agents (described below) are said to know the fundamental value $p^f$ of the 
asset. The value is determined at every timestamp $t$ and assumed to follow a 
random process which causes jumps and subsequently leads to readjustment of the 
agent behaviour 
\begin{equation}
\label{eq:11}
p_t^f =
\begin{cases}
	p_{t-1}^f (1 + j) &\text{if } u_t < \iota \\
	p_{t-1}^f&\text{otherwise}\\
\end{cases}
\end{equation}
\\
where $j$ is an information adjustment constant by which the previous 
fundamental value is updated, $\iota$ is the information threshold which 
determines when new information shall be incorporated in the price and $u_t$  
is drawn from a uniform distribution $\mathcal{U}(0, 1)$. 

\subsection{Agents} \label{sec:Agents}

Four types of agents are modelled to create realistic dynamics, namely 
fundamentalists, chartists, noise traders (hereafter, as a group, also called 
stylised traders or agents) and dealers. While \cite{Chiarella.2008} model a 
representative agent by combining fundamental, chartist and noise trader 
forecasts, this article deals with the different behaviours separately. Hence, 
one can think of it as splitting the representative agent in equation 
(\ref{eq:1}) into its parts. Stylised traders derive their return forecast 
($\hat r$) by applying the following rules:
\vspace{0.4em}
\begin{center}
\begin{equation}
\begin{tabular}{ll}
	fundamentalist &  $\hat r_t^f = \frac{p_t^f-p_t}{p_t} + \eta_t^f$ \\
	chartist       &  $\hat r_t^c = \bar r_{L_i} + \eta_t^c$ \\
	noise trader   &  $\hat r_t^n = \epsilon_t + \eta_t^n$
\end{tabular}
\end{equation}
\end{center}
\vspace{0.8em}
where $\eta_t$ and $\epsilon_t$ follow normal distributions with zero mean 
and $\sigma_{\eta}$ and $\sigma_{\epsilon}$ variances, respectively. 

Further, stylised agents maximise their expected wealth under a CRRA 
utility. To the best of my knowledge this is the first work assuming that 
agents' allocation decisions in a LOB setting depend on their wealth. In 
previously published research, CARA utility functions are applied as it relaxes 
the computational need to keep track every agent's wealth in every simulation 
iteration. However, this approach limits the dynamic nature of investment 
decisions as it decouples the number of shares an agent holds from their wealth.
In contrast, the utility chosen for this study
\begin{equation}
\label{eq:12}
u^i(W, \gamma) = \frac{1}{1-\gamma^i} W^{1-\gamma^i} \text{.}
\end{equation}
allows for a more sophisticated modeling of agent behavior, integrating 
wealth as a key determinant in decision-making. This model is essential 
for accurately representing how agents determine their share allocations, 
providing a realistic context that extends beyond price information, which is  
the single influence in a CARA setting. Furthermore, the selected utility enables 
the implementation of short-selling constraints, allowing stylized agents to short 
sell a multiple of their wealth, which is challenging to incorporate effectively 
under a CARA framework.

Under simplifying assumptions, such as Gaussian returns, the investment 
fraction at timestamp $t$, $Z$, is defined as\footnote{The derivation can be 
found in 
\cite{Chiarella.2001}, Appendix A.1. Note how wealth, comparing to equation 
(\ref{eq:5}), is no longer part of the denominator.}
\begin{equation}
\label{eq:13}
Z_t^i = \frac{ln(\hat p_{t+1}^i / p_t)}{\gamma^i {\sigma_t^{2}}} \text{.}
\end{equation}
with $\sigma_t^2$
being the market variance leaving out the superscript for agent $i$ for better 
readability.

After forecasting the return of the next period and determining their order 
size, stylised agents have to decide about the order price to be transmitted to 
the exchange. Results by \cite{Bouchaud.2002} about fat-tailed distribution 
of volumes and the relatively small fraction of market orders in a LOB are 
incorporated. The latter is to some extent addressed by simply allowing only 
limit orders to be submitted; additionally it simplifies the order matching 
process. To include the distribution of volumes, the order prices $p^{bid}$ and 
$p^{ask}$ are determined by a linear transformation of a random value drawn 
from a log-normal distribution $\zeta$ inspired by \cite{Bartolozzi.2010}
\begin{equation}
\begin{aligned}
\label{eq:14}
p^{bid} = p^{bid, best} - (\zeta -  Q_{0.5}) \\
p^{ask} = p^{ask, best} + (\zeta -  Q_{0.5})
\end{aligned}
\end{equation}
where the index \textit{best} indicates the highest bid and the lowest ask 
price in the market and $Q_{0.5}$ is the median of the log-normal 
distribution.\footnote{The shape of the log-normal is fixed at 0.5 and scale at 
10 following the terminology found here: 
\href{https://docs.scipy.org/doc/scipy/reference/generated/scipy.stats.lognorm.html}
{https://docs.scipy.org/doc/scipy/reference/generated/scipy.stats.lognorm.html}.}

In addition to the stylised agents described above, a monopolistic dealer, in 
the fashion of \cite{Avellaneda.2008}, is introduced who passively provides 
liquidity and earns rebates from the exchange for facilitating transactions. 
While the dealer is incentivised to quote by earning the bid-ask spread, she 
faces inventory and asymmetric information risk. Inventory risk arises from the 
uncertainty of the asset's future value and asymmetric information accounts for 
the risk that a better informed trader takes positions against the liquidity 
provider knowing that the asset bought or sold will move in a specific 
direction.

Note that the \cite{Avellaneda.2008} deploy a CARA utility rather than the 
utility type used here for stylised agents. Introducing CRRA utility to the 
dealer behaviour will complicate the model and reduce tractability. The 
liquidity provider would need to choose the order size at every point in time 
depending on her wealth and subsequently influence market dynamics 
significantly. Ceteris paribus comparisons will become increasingly involved. 
Moving closer to session end, $T-t$ will reduce the spread which in turn 
increases probability to end the session with a balanced inventory. To improve 
model tractability the dealer framework (equations (\ref{eq:7}) and 
(\ref{eq:8})) is adjusted by omitting the time aspect. Ignoring this feature 
allows to reduce an influence which may tilt the result in a particular way and 
complicate the interpretation.

The extension by \cite{Fushimi.2018} already introduced in equations 
(\ref{eq:9}) and (\ref{eq:10}) will also be analysed. The authors introduce a 
simple rule for managing inventory to avoid excess build up. Hence, the 
extension itself is not outcome of the utility maximisation problem, but rather 
a practitioner's heuristic. Agent-based simulations have demonstrated that the authors' extensions do not yield the desired outcomes as claimed when integrated with the reservation price framework. Due to this observation the heuristic will be used under the 
assumption that the reservation price is set to the currently prevailing market 
price, $p_t^r = p_t$. Additionally, the AS model suffers a drawback given the 
relationship between risk aversion and spread remains positive only for relatively large values of $\sigma$. As soon as it is below unity, the spread in equation (\ref{eq:8}) becomes a decreasing function of risk aversion. Given that it is not intended by the authors, I am going to refer to risk aversion as $\frac{2}{\gamma}$ to keep the analogy the same surrendering some interpretability of absolute levels of risk aversion in the following parts. Considering that the agent-based model delivers simulations for high-frequency interactions, the computation of realised-variance leads to 
very small values. Applying those in the model will not lead to meaningful 
results which is why a scaling factor is added to the variance.\footnote{ 
Scaling of $24 * 60$ can be interpreted in such a way that the dealer is 
operating on a 20s timescale in a 8 hour trading session.}

Contrary to most of the published work on financial ABMs, all agents are 
allowed to short-sell. Stylised traders can hold long positions up to the maximum of their wealth and take short positions in an equivalent currency amount. Note, when the price moves up against a short position, the relative investment fraction $Z_t^i$ ones 
instantiated moves below $-1$. As soon as the agent is re-selected to take an 
investment decision, the new wealth is taken into account and the position has 
to be cut. While a dealer as depicted above can also be both long and short, 
she does not face any constraints on the investment fraction.

\subsection{Simulation}

At initiation of every simulation the fundamental price $p^f$ is set to 1,000. 
100 simulations each with 40,000 simulation steps are conducted to be able to 
robustly estimate statistics while stile limiting computational costs.
The simulation involves a population of $N^A = 1,000$ agents of which one is 
a monopolistic dealer. The remaining population is split 
across fundamentalists with $N^f = 450$, chartists $N^c = 450$ and noise 
traders $N^n = 99$. Stylised traders
are randomly assigned an amount of stock $S^i$ and cash $C^i$, both drawn 
from a uniform distribution - $S_0 \in [N_S^-, N_S^+]$, $C_0 \in [N_C^-, 
N_C^+]$. $N_S^- (-1) = N_S^+ = 2,000$, $N_C^- = 2,000$ and $N_C^+ = 10,000$ 
were used across all runs.
The dealer receives no shares to start the simulation with a balanced inventory 
but 
$C_0^{dealer}$ is set to $5,000,000$. Also, following parameters are fixed: 
$\Delta = 
0.1$, $\tau=50$, $\omega = 0.1$, $\sigma^f = \sigma^n = 0.0005$, $\sigma^f = 
0.001$, $\gamma = 0.1$ for the dealer and $\gamma=10$ for 
stylised agents, $\kappa = 0.6$, $\phi^{bid} = \phi^{ask} = 5,000$ and 
$\eta^{bid} = \eta^{ask} = \frac{log(1 / \phi)}{\phi}$. The behavioural rules 
of all agents take some form of variance into account which will be computed in 
all circumstances as an exponential moving average with decay $\alpha = 0.25$.

At every timestamp $t$ either dealer or stylised traders can place their 
orders. If stylised traders are selected, one agent is randomly drawn from the 
population. Note, if the dealer is selected, she can place up to two orders 
and, hence, consumes two consecutive timestamps. All other agents can only 
place one order at a time. After the order submission of the dealer, only the 
stylised traders can place their order.

\section{Results}

This section presents the results of an agent-based simulation on how liquidity 
provision affects market behaviour. I begin with defining a set of financial 
objectives and parameters to guide the analysis. Afterwards a contextual 
transition from probabilistic simulations to agent-based modelling in the 
context of the earlier defined dealers and comparison of their performance in 
the main test environment, optimal behaviour and preferences of strategies are 
discussed. Lastly, the derived optimal behaviour is put into the context of 
impact on market dynamics and stylised agents.

\subsection{Financial Objectives of Liquidity Providers}

\cite{Avellaneda.2008} developed a model for optimal liquidity provision that 
is primarily concerned with inventory risk management. Their derivation uses an 
expected utility framework which creates a dependence between dealer quoting 
(indirectly influencing the order flow) and wealth. As such, in a probabilistic 
environment the dealer is always able to manage her inventory position and 
maintain a balance. However, financial markets are raising the bar as they do 
not necessary follow probabilistic assumptions made. To better capture observed 
financial market behaviour in an agent-based simulation I summerise the dealer 
objectives as follows.

\begin{enumerate}
\item \textbf{Performance objectives}. As in \cite{Avellaneda.2008} the dealer 
maximises her expected utility of wealth and does so by increasing her profits. 
This is the goal of any market participant, however, the difference lies in the source of 
wealth gain. While majority of investors try to harvest some sort of risk 
premia, a dealer - especially a liquidity provider - is primarily focused on capturing the bid-ask spread along with rebates for quoting. These returns are expected to be relatively stable in nature as they are unaffected by trends in the underlying.

\item \textbf{Risk objectives}. To achieve maximum utility, the dealer will 
adjust her quotes to shift the chances of order flow in her favour. The idea 
is to mitigate inventory risk by keeping it as neutral as possible to avoid 
being impacted by uncertainty about the underlying price. Put differently, 
accumulating large inventories has to be avoided, also for the reason to not 
become a substantial holder of the underlying and thereby turn out to be 
systemic for the whole market. By introducing \cite{Fushimi.2018} adjustment, 
in addition to explicitly avoiding accumulation of large positions one can also 
limit the risk of adverse selection by setting a cap for the maximum order size.
 
\end{enumerate}

\subsection{The Importance of Inventory Risk Management in Agent-Based Limit 
Orderbooks}

The results that follow are based on agent-based simulations of the baseline 
models to introduce the behaviour of the different type of dealers described 
earlier. The goal is to highlight the importance of managing inventory, so a 
naive liquidity provider is used as a benchmark. In comparison to the 
probabilistic simulations in section \ref{sec:Inventory 
Risk} the naive dealer is able to generate significant profits from her 
activities. She experiences an average total return on wealth of 30.0\% 
comparing it with the AS and IR model of 25.3\% and only 1.5\%, respectively. 
However, the wealth evolution path of the naive model is materially more risky 
with an average volatility of 48.9\% versus only 2.7\%/1.3\% of the AS/IR 
model. Risk-adjusted performance clearly favours the strategies that explicitly 
controls inventory. Strikingly, the performance dominance has now changed in 
favour of the AS dealer as opposed to the probabilistic simulation where the IR 
dealer was the clear winner. 

\begin{figure}[!htbp]
\centerline{\includegraphics[width=6in]{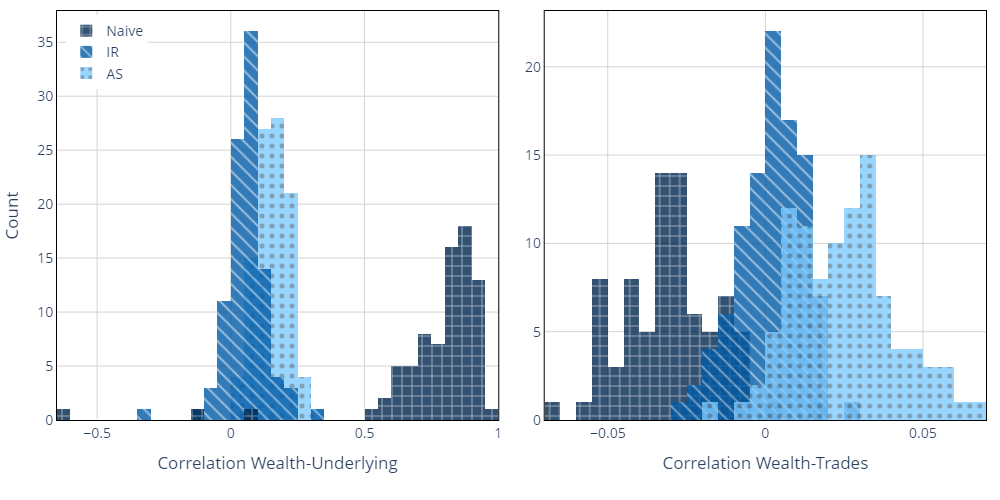}}
  \caption{Histogram of correlations for all simulation runs.} 
  \label{fig:histogram_2}
\end{figure}

To draw a full picture of the comparison it is important to understand why one 
can see such a large difference in total return and volatility. When taking a 
closer look at the evolution path of the dealer's wealth and the underlying, it 
becomes obvious that the naive dealer's wealth is highly correlated to the 
underlying. The reason becomes very apparent as the dealer accumulates large 
inventories over time without the possibility to get rid of it. Hence, her 
wealth becomes highly dependent on the movement of the stock. A positive 
correlation but to a lesser extent is also visible for the AS dealer. The 
origin is similarly related to the accumulation of stock: While being able 
to keep the inventory in balance, she is on average exposed to a positive 
inventory and as such positively correlated to the stock market.
The left panel of Figure \ref{fig:histogram_2} contains a histogram of the 
wealth and underlying correlation across all simulation runs and visually shows 
these findings. The right panel visualises the correlation between wealth and 
trade value. While the naive strategy holds significant amounts of stocks, her 
trades are slightly negatively correlated to her wealth, a pattern consistent 
with purely passive quoting. Simply said, the market price has to move down to 
her bid (and vice versa) to create the inverse association. The other two 
strategies show no meaningful correlation different from zero.

In attempting to reconcile the results with the financial objectives defined in 
section 5.1, it becomes apparent that neither objective is satisfied by the 
naive model. While the dealer is able to generate significant profits, her 
returns are inferior on a risk-adjusted basis benchmarked with an AS dealer. 
Moreover, neutrality towards trends in the underlying is not achieved, but even 
worse so, the naive dealer becomes a systemic risk through accumulation of 
stocks. The strength of the AS model emerges very tangibly by ticking the box 
of a majority of the objectives. Risk-adjusted returns are high and the dealer 
does not influence the market with her inventory, but a positive correlation to 
the underlying remains and it is yet to be understood how receptive the 
accumulation of stock is towards parameter sensitivity. The IR model clearly 
lacks performance and trades off risk adjusted returns for strictly managing 
inventory. The performance difference to the probabilistic simulation which 
favoured the AR model is sizeable and demonstrates the need for a more 
realistic simulation environment. Thus, the next sections will delve into 
greater detail on the sensitivities of the dealer's parametrisation.

\subsection{Optimal Behaviour of Liquidity Providers}
In this section the optimal behaviour of liquidity providers will be analysed 
by simulating different preference specifications. In particular, I study how 
risk aversion preferences, maximum order size limitations and inventory 
skewness (IR model only) influence optimal behaviour.

Risk aversion influences the dealer's 
behaviour in two distinct ways: 
\begin{enumerate}[label=(\roman*)]
	\item Higher risk aversion tilts the reservation price away from the 
	inventory side. To be precise, assuming the dealer has a long 
	position in the stock, a higher risk aversion will lower the reservation 
	price and vice versa. The dealer is offering to sell down her 
	inventory at a lower price than traded in the market or is willing to 
	collect more stocks but only at a discount. The same logic can be applied 
	to a short position in the asset.\footnote{Note that this applies only for 
	the AS model.}
	\item The quoted bid-ask spread around the reservation price is wider for 
	higher risk aversion. This implies a more conservative stance of the dealer 
	as she requires a larger premium for providing liquidity.
\end{enumerate}

\begin{figure}[!htbp]
\centerline{\includegraphics[width=6.5in]{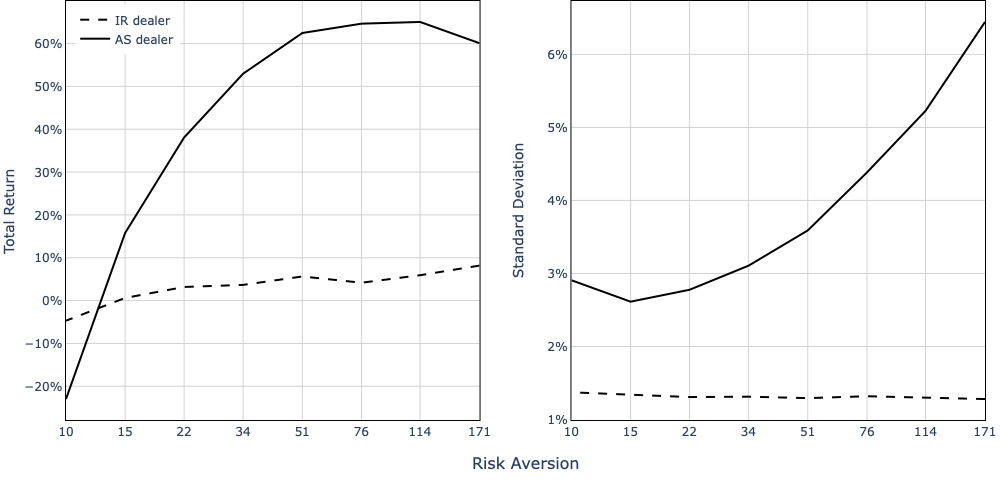}}
  \caption{The sensitivity of wealth return statistics such as standard 
  deviation and total return to changes in the dealer's risk aversion. Note 
  that the x-axis displays risk aversion as $\frac{2}{\gamma}$.}
  \label{fig:ra_mm_wealth_1}
\end{figure}

As can be seen in Figure \ref{fig:ra_mm_wealth_1}, both type of dealers 
experience higher 
wealth growth at increasing levels of risk aversion. A significant gap opens in 
favour of the AS model which outperforms the IR model by large margins. 
However, while favouring higher levels of risk aversion, the AS dealer's 
improvement experiences diminishing marginal returns. In contrast, the IR model 
has a seemingly linear effect in improvement. 
The reason lies in the statistical properties of the order book: Given that the 
quotes are set wider, the probability of being executed on the resting orders 
becomes increasingly low, which in turn reduces the trading frequency and so 
the number of opportunities one can capitalise on. 

Contrasting, lowering risk 
aversion comes at an increasing downside. This is aggravated by the 
fact that quotes will be at the top-of-the-book most of the time and thus 
putting chances high that taking the trade with the stylised traders too early 
at prices too close to the mid. From a different perspective, the premium one 
can gain from the bid-ask spread narrows through self-inflicted behaviour of 
the dealer. These effects are intensified by the constant order size of the AS 
model, while the IR model can mitigate them via the prudent inventory 
management rules. This is visible in the lower threshold that both strategies 
show at the lower end of the parameter range where the AS dealer starts to 
significantly underperform its counterpart.

Considering the variance of their returns and taking a closer look into the 
trading volume, it becomes obvious that the AS model outperforms because it is 
able to trade and hold a higher amount of shares. While the average trade size 
is twice as large for the AS model compared to the AR model, the average 
inventory grows exponentially 
supporting this fact. 
On a risk-adjusted basis, the IR model can reach fairly high Sharpe ratios at 
the maximum parameter range which is supported by the unchanging volatility.
The AS agent requires fairly low risk aversion levels to maximise risk-adjusted 
returns and will not see any benefits of further increases due to compensating 
effects of volatility which originates from longer holding periods and larger 
inventory of stocks. Higher order moments like skewness and kurtosis benefit 
from increasing risk aversion. Allowing the IR agent to adjust her risk 
aversion leads to a significantly better achievement of the performance 
objectives compared to the baseline but is still outperformed by the AS model. 

\begin{figure}[!htbp]
\centerline{\includegraphics[width=6in]{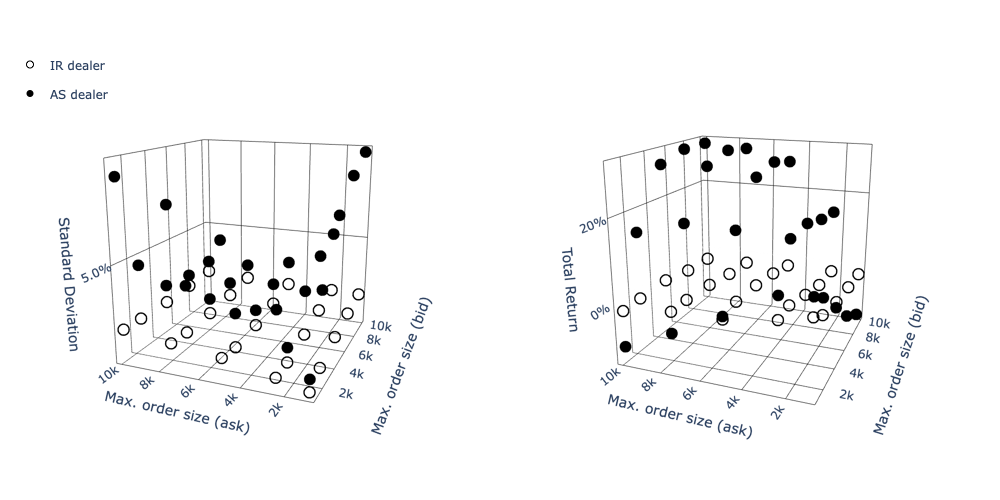}}
  \caption{The sensitivity of wealth return statistics such as standard 
  deviation and total return to changes in the dealer's maximum order size 
  $\phi^{max}$.} 
  \label{fig:mos_mm_wealth_1}
\end{figure}

Exploring the sensitivity towards the maximum order size, asymmetric effects 
emerge. I observe a similar pattern in adjusting maximum order size as I did 
for changes in risk aversion: The AS dealer experiences diminishing returns for 
increasing order size. The results are visualised in Figure 
\ref{fig:mos_mm_wealth_1}. Strikingly, a very 
asymmetric quoting, e.g. low ask (bid) quote and high bid (ask) quote leads to 
negative returns, also lower than the IR dealer. This is related to the fact 
that when building up stock position with a larger bid size, the agent cannot 
balance back the inventory similarly fast with a lower ask size and vice versa. 
It is supported by the significantly higher volatility of wealth returns at a 
high degree of quote asymmetry and an average inventory that is tilted to the 
positive side for larger bid size. Interestingly the skewness increases for the 
asymmetric quoting and kurtosis sinks. The latter is attributable to the 
increased variance of the inventory position which results in a higher wealth 
volatility and as such creates a wider distribution. The effects on skewness do not seem to follow a fundamental reason and as such are not possible to explain.
To conclude on asymmetric quote sizes, both dealer experience a significant increase in magnitude of correlation of their wealth to the underlying towards unity, both positive and negative depending on the tilt of the quote size. This is introduced by the accumulation of 
inventory that is not easily balanced off any more and thus is also sub-optimal 
from the risk objectives perspective. 

While the IR dealer\footnote{Given that the inventory skew has significant 
effects on the order size, it is adjusted for every parametrisation in such way 
that order size goes to unity at max. order size.} sees impacts of total return 
on a much smaller scale, she prefers lower order sizes and does not have any
asymmetric effects. Additionally, her volatility of wealth returns is much more 
contained but she experiences worse 3rd and 4th moments overall. Surprisingly 
both dealers experience shrinking correlation of wealth and underlying return 
with increasing (symmetric) order size in line with better risk objectives. The 
correlation for the AS dealer sinks from 0.25 towards 0.15 whereas the IR 
dealer sees a drop from 0.1 to 0.05. While not fully explainable it appears to 
be related to the increasing correlation of wealth and trades, which rise 
towards zero from low negative values. 

\begin{figure}[!htbp]
\centerline{\includegraphics[width=6.5in]{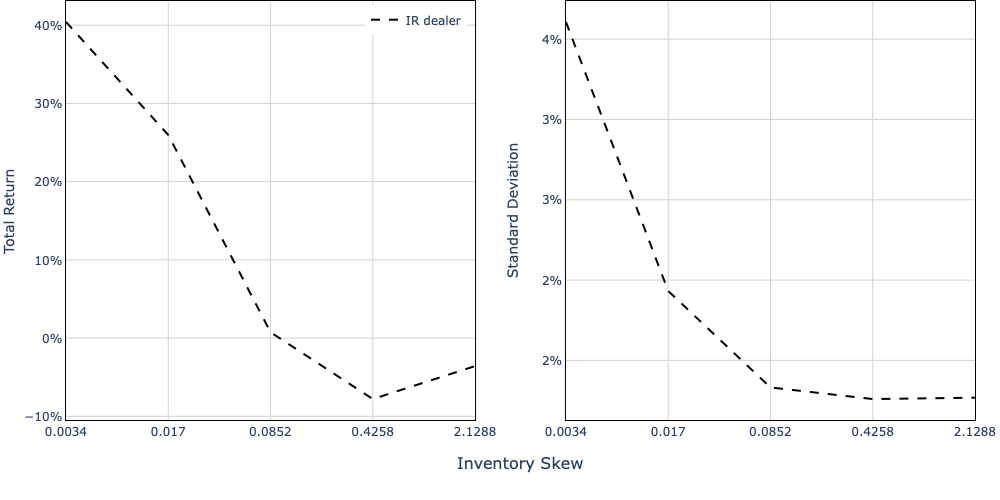}}
  \caption{The sensitivity of wealth return statistics such as standard 
  deviation and total return to changes in the dealer's inventory skew 
  $\eta$ in absolute terms. The graph shows symmetric inventory skew for bid 
  and ask. Note that parameters are scaled by factor 100 for better 
  readability and are displayed on a log-scale.} 
  \label{fig:is_mm_wealth_1}
\end{figure}

As inventory skew is only relevant to our IR agent, this part cannot be 
compared to the AS model.\footnote{See Figure \ref{fig:inventory_skew_demo_1} 
in 
Appendix \ref{appendix:a} to make the argumentation more tactile.} When 
lowering the inventory skew towards zero the quote size will approach the 
maximum ($\phi^{max}$) that is also used by the AS dealer. It is visible in 
Figure \ref{fig:is_mm_wealth_1} that the total return increases significantly 
and comes close to ranks achieved by the AS dealer at 
lowest skew levels. This is accomplished by removing restrictions on the 
inventory accumulation and rather letting it be dictated by the market. That is 
also the reason why correlation of wealth to the underlying increases at the 
lowest specification simulated. Contrary to the AS model, the IR dealer does 
not tilt her reservation price but rather keeps it constant and thus does not 
alter the probability of keeping inventory as neutral as possible. This is 
supported by a large increase of the average inventory size and volatility, 
which rises to nearly three-fold of the base specification. 
A slightly lower skew than in the base specification however leads also to a 
large increase in total returns (to about 26\%) without the negative side 
effects just mentioned and by that maximises risk-adjusted returns. The 
strategy is able also to perform closely to the AS model on a risk adjusted 
basis. Similarly to the already analysed asymmetric max. order size, asymmetric 
inventory skew results in accumulation of stock, along with it a high 
correlation to the underlying and, thus, can be rejected from the optimality 
discussion.

Both dealer strategies show great differences in their wealth moments and as 
such possess partly contrasting preferences towards their parameter 
specification. Ultimately both AS and IR dealer are able to meet financial 
performance as well as risk objectives and at the same time also optimise the 
their relevant metrics. While both dealer prefer higher risk aversion, the IR 
model prefers the highest specification in terms of risk-adjusted returns 
whereas the AS model does best for low-to-middle risk aversion levels. Both 
models show congruent distaste for asymmetric order size. In regard to 
symmetric quoting, the IR model prefers lower order size with which she comes 
close to the Sharpe ratios gained by the AS dealer. The AS dealer, here 
again, sees diminishing risk-adjusted returns and can already reap all benefits 
with max. order size of medium range (at around 5000 shares). The IR dealer can 
adjust her inventory skew towards the lower tested spectrum to increase the 
total return significantly. Going to the lowest inventory skew results in a 
similar behaviour as in the naive strategy since the reference price is not 
tilted to balance the inventory.

\subsection{The Effects on Price Discovery}
After carefully examining the optimal behaviour of the dealers, it is now time 
to understand their impact on market dynamics and other agents.

\begin{figure}[!htbp]
\centerline{\includegraphics[width=6.5in]{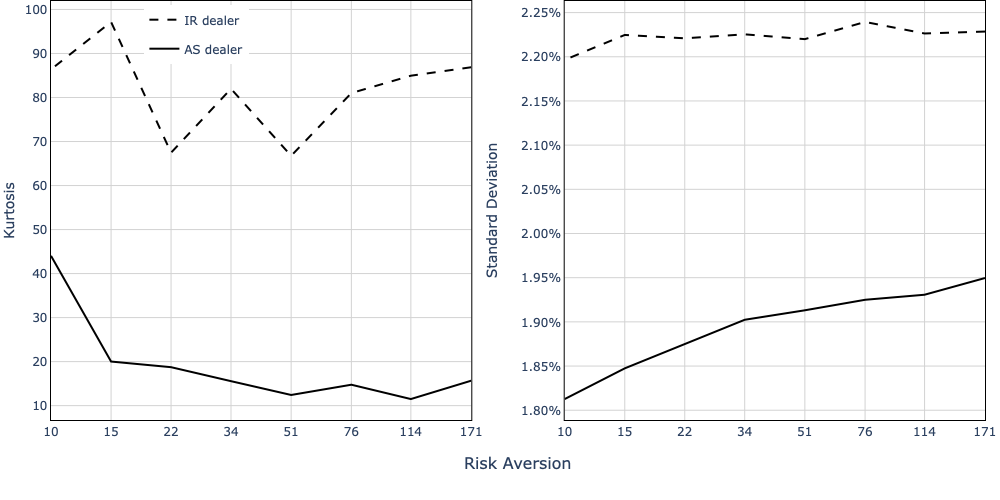}}
  \caption{The sensitivity of market return statistics such as kurtosis and 
  standard deviation to changes in the dealer's risk aversion. Note that 
  the x-axis displays risk aversion as $\frac{2}{\gamma}$.} 
  \label{fig:ra_market_1}
\end{figure}

While being an important determinant in personal wealth for both dealers, risk 
aversion in the IR model has no significant effects on market dynamics. 
However for the AS model, higher risk aversion results in greater market 
volatility and a lower return kurtosis as shown in Figure 
\ref{fig:ra_market_1}. 
The lower return kurtosis is a direct consequence of the increased volatility 
which in turn stems from a larger bid-ask spread of the dealer so that the 
price has more room to fluctuate. If an order arrives to 
the market, it  will be first matched at price priorities and if large enough 
it will be eventually matched with the dealer. However, in such a situation it 
is highly unlikely that it will not be filled completely and matched against 
the quotes further away from the mid. It can be thought as if the dealer sets a 
hard stop for the price to fluctuate above or below her quotes. From a 
different perspective, the likelihood of hitting a dealer order with lower risk 
aversion is increased along with the likelihood that the order ripples through 
the LOB and as such increases the tails. Notably, a market with an AS dealer 
experiences a generally lower level of kurtosis and volatility as compared to 
the IR dealer.

Similarly, stylised agents are not affected by the IR strategy but show some 
effects in the AS model. Here, fundamentalists are the major beneficiaries and 
can increase their total return. This can be explained by their superior 
information regarding the fundamental price which they can exploit more easily 
given a lower bid-ask spread at low risk aversion.

\begin{figure}[!htbp]
\centerline{\includegraphics[width=6.5in]{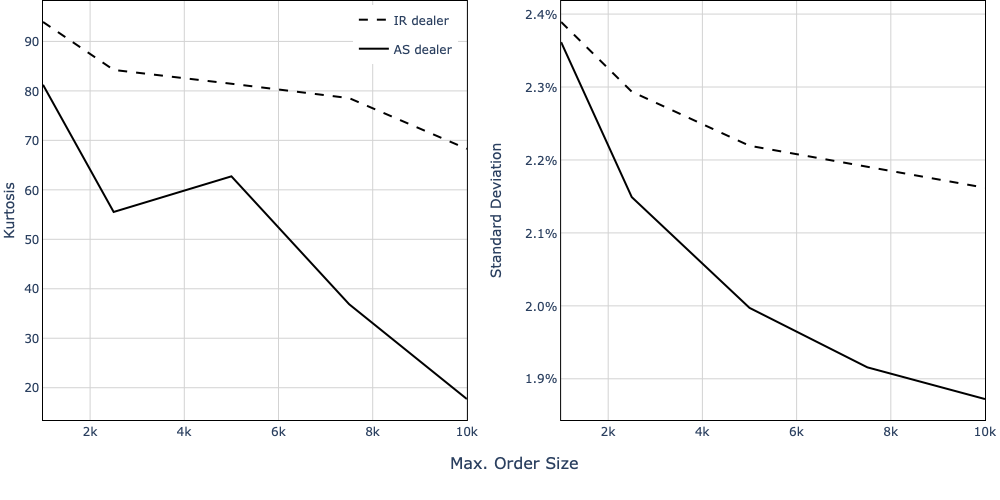}}
  \caption{The sensitivity of market return statistics such as kurtosis and 
    standard deviation to changes in the dealer's max. order size.} 
  \label{fig:mos_market_1}
\end{figure}

For both dealers, a larger order size lowers market volatility, as shown in 
Figure \ref{fig:mos_market_1}. Contrary to the findings for risk aversion, a 
larger order size also lowers the market returns' kurtosis.
The principle mechanism is that a higher order size acts like a roadblock for 
incoming orders, as described above. The larger the size, the smaller the 
probability that an order can pass it. Comparably, also stylised agents benefit 
from a larger max. order size. Both, fundamentalist and chartist exhibit higher 
total returns, lower volatility and kurtosis of wealth which results in 
superior Sharpe ratios. The effects of larger order size are more pronounced in 
markets operated with an AS dealer.

As expected from a lower inventory skew it lowers market volatility around 20\% 
and is established through larger dealer quotes and 
longer time in force. Through the lower skew, the size of the 
quote is less drastically reduced and, consequently, translates into limit 
orders being longer in the market. Note that if saturated, the IR dealer can 
even reduce the quote  
to zero, which is less likely the lower the inventory skew. The influence on 
market kurtosis is unintuitive at first as it peaks at the base 
specification and drops by 35\% and 58\%, both, in the directions of larger and 
lower inventory skew. While the drop in kurtosis towards lower skewness is 
related to larger absorption rate of trades of stylised agents and leads 
to a similar effect as increasing order size, the impact from higher 
skewness seems to stem from the larger variance which in turn lowers the 
kurtosis of market returns.
Lower market volatility also translates into stylised agent wealth. While the 
chartists benefit from higher risk adjusted returns at lower inventory skew, 
which are both driven by lower volatility and higher total returns, 
fundamentalists' Sharpe ratios suffer from it. They are favoured by higher 
skewness but experience diminishing marginal risk-adjusted returns which are 
already maximised by the base specification. The additional return gained 
through higher skew is offset by the larger volatility. However, all stylised 
agents experience negative wealth evolution here as well.

Putting preferences and market impact together, both mutually beneficial but 
also detrimental behaviour can be derived. As laid out in the previous section, 
the optimum from a risk-adjusted return perspective in the AS model is in the 
low-to-mid range of risk aversions, and, hence does not lower volatility as 
much as it can but reduces kurtosis. Here, the IR strategy's impact on 
the market is comparably small. The question about optimality from 
an objective perspective cannot be answered as the preferences from 
other market participants are not known. In particular this relates to the 
trade-off whether agents prefer lower volatility or kurtosis.
The AS and IR models show opposite effects on their wealth concerning the order 
size. While the later prefers lower order size the former favours 
larger size, 
but comes at a diminishing return. Markets clearly honour larger quote size 
as is apparent in both lower kurtosis and volatility. This results in 
market-aligned behaviour of AS dealer while the IR model's impact is generally 
negative.  
Lower inventory skew imposes positive effects on the market by both lowering 
volatility and kurtosis along with being preferred by the IR dealer who can 
increase her performance significantly. The effects on stylised agents are 
ambiguous as chartists clearly benefit from lower inventory skew while 
fundamentalists suffer and vice versa. For the IR model, the total effect is 
somewhat compensating as the dealer would favour lower order size and lower 
skewness. As just mentioned, the latter would be beneficial for market dynamics 
while the first would be partially counteracted.

\section{Conclusion}
This paper developed an agent-based model which centres around a limit orderbook
allowing market participants to share information and thereby lead price 
discovery. Market participants are three types of stylised agents - 
fundamentalists, chartists and noise traders - additionally I introduce passive 
dealers who act as liquidity providers.
The importance of agent-based modelling for trading simulation is shown by 
providing evidence of different conclusions depending on which framework to use 
as a test bed. The probabilistic model neglects dynamic interactions and 
endogenous behaviour that are an important part of real financial markets. As 
such, evidence was presented that naive dealer strategies accumulate large 
amount inventory and experience large correlations of their wealth to the 
underlying, which both could be found through the agent-based framework. The 
extension by \cite{Fushimi.2018} (IR dealer) showed much lower performance when 
deploying it in an agent-based simulation and was outperformed by the model 
developed in \cite{Avellaneda.2008} (AS dealer). 

Both dealer models have been further studied to understand their preferences 
and subsequent effects on market dynamics. Dealers have commonalities by 
preferring higher risk aversion as it significantly impacts their wealth. 
However, this has ambiguous effects on the market which is visible in higher 
return volatility but shallower tails. With respect to max. order size, the AS 
model benefits from larger quotes which was shown to be in the interest of 
markets by reducing both volatility and kurtosis. By comparison, the IR dealer 
prefers lower quote size and would in fact impact markets negatively. Inventory 
skewness has large implications for the dealer's wealth as she prefers to lower 
her quote size moderately and, thus, also has positive impact market dynamics. 
At the same time chartist benefit but fundamentalists are disadvantaged, both 
of which is visible in their total returns. The analysis has shown that agent's 
returns clearly depends on the dealer's implementation what her preferences are 
and how those loop back to market dynamics.

This work uncovered some important relationships between 
preference parameters on market dynamics and wealth evolution. This framework 
can be extended to include other variables by incorporating completely new 
rules or environments similar to the work done by \cite{Vuorenmaa.2014}. 
However, the model should remain tractable when introducing significant 
extensions. Future work can include more realistic behaviour of stylised agents 
to make them profitable or a more active market maker strategy. 

\clearpage

\appendix

\section{}
\label{appendix:a}

\begin{figure}[!htbp]
\centerline{\includegraphics[width=7in]{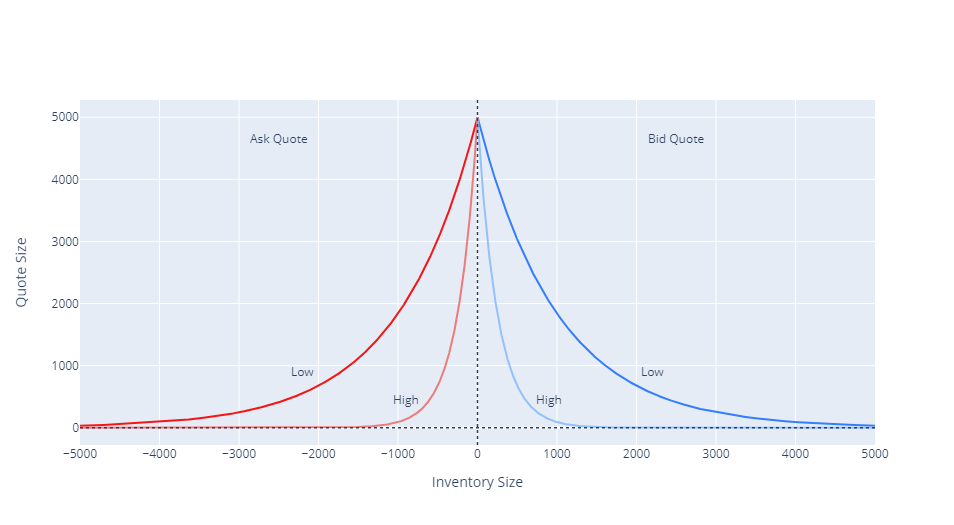}}
  \caption{Demonstration of effects of low and high inventory skew on quote size given different 
  values of inventory size. High (low) inventory skew leads to faster (slower) decay of quoted volume 
  in the direction inventory. Maximum order size is set at 5000, low (high) skew demonstrates a 
  parameter value of 0.001 (0.004).} 
  \label{fig:inventory_skew_demo_1}
\end{figure}

\clearpage

%
%
%

\clearpage


\bibliographystyle{jfe}
\bibliography{abm}

\end{document}